\documentstyle[twocolumn,prl,aps]{revtex}
\begin{document}
\twocolumn[\hsize\textwidth\columnwidth\hsize\csname
  @twocolumnfalse\endcsname
\title{Theory of Incompressible States in a Narrow Channel}
\author{Tapash Chakraborty}
\address{Institute of Mathematical Sciences, Taramani, Madras
600 113, India}
\author{K. Niemel\"a and P. Pietil\"ainen}
\address{Theoretical Physics, University of Oulu,
Linnanmaa, FIN-90570 Oulu, Finland}
\date{December 2, 1996}
\maketitle
\begin{abstract}
We report on the properties of a system of interacting electrons 
in a narrow channel in the quantum Hall effect regime. It is 
shown that an increase in the strength of the Coulomb interaction
causes abrupt changes in the width of the charge-density profile
of translationally invariant states. We derive a phase diagram 
which includes many of the stable odd-denominator states as well 
as a novel fractional quantum Hall state at lowest half-filled 
Landau level. The collective mode evaluated at the half-filled 
case is strikingly similar to that for an odd-denominator 
fractional quantum Hall state.

\end{abstract}
  \vskip 2pc ] 

\narrowtext
The existence of incompressible quantum fluid states in a
two-dimensional electron system subjected to a strong
perpendicular magnetic field has presented us with a profound
understanding of the odd-denominator fractional quantum Hall 
effect (FQHE) \cite{fqhe,laughlin,halperin,book}. Interestingly, 
such a clear physical understanding of the simplest 
even-denominator state, viz., the half-filled lowest Landau 
level, has not yet emerged \cite{halperin,expt,theory,electron}.
In recent years, study of electron correlations in narrow 
channels has received increasing attention \cite{okiji}. 
Observation of a new incompressible state at half-filled Landau 
level in a narrow channel was indeed reported a few years ago 
\cite{timp}, and is naturally expected in the Laughlin picture
generalized to one-dimension \cite{chui}, where the statistics
of the charge carriers should be arbitary \cite{frad}. That 
observation was remarkable because such a state has, as yet, not 
been found to appear, either in experiments \cite{fqhe,expt} or 
in numerical studies \cite{electron} of a two-dimensional 
electron gas. There have been several attempts to explain the 
origin of the nonexistence of a stable half-filled quantum Hall 
state in two-dimensions. These include, among other things, a 
transformation from electrons to fermions with a Chern-Simons 
field \cite{theory}. One other possible explanation was suggested
in Ref. \cite{electron,greiter} where it was shown that reduction
of the short-range part of the repulsive electron-electron 
interaction is required to stabilize this state.

Here we report on the results of a model we have developed for 
the FQHE in a narrow channel (1D-FQHE) where there are a finite 
number of spin polarized electrons subjected to a strong
perpendicular magnetic field, interacting via the long-range
Coulomb potential and confined by a potential, which is 
{\it parabolic} in one dimension and flat in the other. 
The electrons are considered to be in a cell 
whose length in $x$-direction is denoted by $a$. Width of the 
cell depends on the strength of the confining potential relative 
to the strength of the interactions and also on the length of the
cell. Imposing periodicity condition in the $x$-direction, the 
system models an infinitely long quantum wire. Our novel results
in this system include abrupt changes in the electron density
profile as one moves from one FQH state to another. These stable
states, which also include the unique lowest half-filled Landau 
level, are identified from their gap structures in the excitation
spectra.

The total Hamiltonian for the system is
\begin{equation}
        {\cal H}={\cal H}_0+{\cal H}_{\rm int},
\label{Hamiltonian}
\end{equation}
where ${\cal H}_0$ contains the kinetic energy of $N$ electrons
of mass $m^*$ and the electrostatic confining potential
\begin{equation}
  {\cal H}_0=\sum_{i=1}^N\left[\frac1{2m^*}\left({\bf p}_i-
  e{\bf A}_i \right)^2+\frac12 m^*\omega_0^2y^2_i\right],
\label{single-electron}
\end{equation}
and ${\bf A}$ is vector potential in {\it Landau gauge}. The 
interaction term of the Hamiltonian consists of the Coulomb 
repulsion of the electrons, the electrostatic energy of the 
positive background, and the interaction energy between the 
background and the electrons.

The single-electron wave functions are given by
\begin{equation}
\psi_\kappa({\bf r})=\left(\frac1{a\sqrt{\pi}\lambda}
     \right)^\frac12
     \exp\left(ikx-\frac{\hat{y}^2}{2\lambda^2}\right)
  H_n\left(\frac{\hat{y}}{\lambda}\right),
\label{spwf}
\end{equation}
where the magnetic length is defined as
$\lambda=\left({\hbar}/{m^*\Omega}\right)^{1/2}$, and
$\Omega=\left(\omega_0^2+\omega_c^2\right)^{1/2}$, where 
$\omega_c=eB/m^*$ is the cyclotron frequency, $\kappa=\left\{n,m
\right\}$, and
$$\hat{y}=y+\frac{\hbar\omega_c}{m^*\Omega^2}k=y+\frac{2\pi
\lambda^2}{\gamma a} m,$$
with a dimensionless quantity $\gamma=\sqrt{1+\left(\omega_0/ 
\omega_c\right)^2}$. In (\ref{spwf}), $H_n$ is a Hermite 
polynomial of order $n$. Along the wire the wave function 
(\ref{spwf}) is just a plane wave with wave vector $k=(2\pi/a)m$.
Here $m$ stands for momentum in the direction of the wire. In the
lateral $y$-direction the wave function has a Gaussian form. 
Restricting ourselves in the lowest Landau level, $i.e.$ setting
$n=0$, and ignoring the constant Landau level energy, the 
single-electron Hamiltonian  (\ref{single-electron}) in second
quantized  form is
$${\cal H}_0=\sum_i\frac{\hbar^2 k_i^2}{2 m^*}
        \frac{\omega_0^2}{\Omega^2} a_i^\dagger a_i
        =\sum_i{\cal E}_i a_i^\dagger a_i$$
where $a_i^\dagger$ ($a_i$) is the creation (annihilation) 
operator of a state $i$.

In the non-interacting ground state, $N$ electrons occupy the 
lowest $N$ available single-particle levels. It is reasonable 
to require that the electron density in that state is symmetric 
around the $y=0$ axis, $i.e.$, the total momentum $M=\sum_j 
m_j=0$. This symmetry condition holds for {\it odd} number of 
electrons if $m$ is an {\it integer}, and for {\it even} number 
of electrons if $m$ is a {\it half-odd integer}. Thus, for odd 
number of electrons we have {\it periodic} boundary conditions 
along the wire, and {\it antiperiodic} boundary conditions for 
even number of electrons. When inter-electron interactions are
introduced in the system the electrons start to avoid each 
other. As interactions increase with respect to the kinetic 
energy electrons begin to occupy also higher levels in order to 
reduce their mutual repulsion. Consequently, states other than 
$M=0$ are also realized as a ground state. However, if the ground
state has $M\neq0$, the system is not expected to be in a 
fractional quantum Hall state \cite{book}.

The Coulomb matrix elements in the present model are obtained
from
\begin{eqnarray}
 \lefteqn{{\cal A}_{m_1,m_2,m_3,m_4}}
             \nonumber \\
        & = & \frac12 \int d{\bf r}_1 d{\bf r}_2
             \psi_{m_1}^*({\bf r}_1) \psi_{m_2}^*({\bf r}_2) 
                         v({\bf r})
             \psi_{m_3}({\bf r}_2) \psi_{m_4}({\bf r}_1)
              \nonumber \\
        & = & \frac12 \frac{e^2}{\epsilon\lambda} 
              \exp\left[-\frac12\left(\frac{2\pi}{\gamma a}
              \right)^2 (m_1-m_4)^2\right]
              \nonumber \\
        & \times & \int dq_y'\frac{\exp\left[i
        2\pi(m_3 - m_1)q_y'\right]
        \exp\left[-\frac12\left(\gamma a q_y'\right)^2
        \right]}{\sqrt{\left[
        \frac{2\pi(m_1 - m_4)}{\gamma a}\right]^2 +
        \left(aq_y'\right)^2}}\nonumber \\
        &\times& \delta_{m_1+m_2,m_3+m_4}
\label{calA}
\end{eqnarray}
where the length is measured in units of $\lambda$, ${\cal E}_c=
e^2/\epsilon\lambda$ gives a measure of the interaction
energy, and the dimensionless integration variable is $q'_y=q_y
\lambda^2/(\gamma a)$. It is to be noted that, at this stage
{\it all} possible combinations of the quantum numbers 
$m_1,m_2,m_3$ and $m_4$, which satisfy the law of conservation of
the momentum, are allowed. Clearly, the dominant term is the
one with $m_1$ close to $m_4$. In the case of $m_1=m_4$, however,
the integral in the second term of (\ref {calA}) does not
converge due to the long-range nature of the Coulomb potential. 
To cancel out this divergence we have two choices: We can either 
use a truncated  Coulomb potential \cite{yoshioka1} or neutralize
the system by embedding the wire into a positively charged 
background. We prefer the latter procedure because then the 
long-range effects of the Colomb force are included in our 
calculations. 

Let us first examine how the translationally invariant state,
$i.e.$, the $M=0$ state, changes when we change 
the strength of the interactions with respect to kinetic and
potential energies of the electrons, {\it i.e.}, ${\cal E}_c
/E_0$ (where $E_0=(\hbar^2/2m^*\lambda^2)(\omega_0^2/\Omega^2)$ 
is the energy unit) and the length of the cell $a$. As we vary 
${\cal E}_c/E_0$ while keeping $a$ fixed, the expectation values 
of the kinetic and potential energies change very abruptly from 
one value to another. As the calculation is repeated for other 
fixed values of $a$ we obtain Fig. 1 (a) and Fig. 1 (b) for 
$\langle{\cal H}_0\rangle$ and $\langle{\cal H}_{\rm int}
\rangle$, respectively. The expectation values show rich 
structures in the parameter space spanned by $a=5,\cdots,12.4$ 
and ${\cal E}_c/E_0=0, \cdots,80$. The two energies $\langle
{\cal H}_0\rangle$ and $\langle{\cal H}_{\rm int}\rangle$ jump 
in opposite directions, and therefore the net change in total 
energy does not clearly show the sudden changes in the $M=0$ 
state. However, for a much longer system (at a fixed linear 
density) we expect sharper first-order transitions between the 
different phases. 

As the jump occurs in the parameter space spanned by ${\cal E}_c
/E_0$ and $a$, it indicates a change in the $M=0$ state. One 
earlier work identified the filling factors ($\nu=N/N_s$ where 
$N_s$ is the Landau level degeneracy) $\nu=\frac23$, $\nu=
\frac13$ and $\nu=\frac15$ FQHE states in a system of six 
electrons interacting via a truncated Coulomb potential and 
calculating the overlap with the Laughlin-like wave functions 
\cite{yoshioka1}. These states are also realized in our system 
with real long-range Coulomb potential. In Fig. 1 (a) the 
plateaus corresponding to states at $\nu=\frac13, \frac23, 
\frac25, \frac27$, and $\nu=1$ are indicated by 
arrows. The state at $\nu=\frac13$ can also be characterized by 
calculating the overlap between the Coulomb-$\frac13$ state and 
Haldane's pseudopotential-$\frac13$ state \cite{yoshioka2}. We 
have checked this overlap in our present system and found it to 
vary between the values 0.83 and 0.89 at $a=9.5$.
 
In our quest for a stable half-filled Landau level, we are 
particularly interested to know what happens in between the well 
established FQHE states. For example, what are the states 
realized in between the FQHE states $\nu=\frac23$ and 
$\nu=\frac13$? In this region there are clear jumps in both  
$\langle{\cal H}_0\rangle$ and $\langle{\cal H}_{\rm int}
\rangle$. To get further insight on the $M=0$ states realized in 
the wire, we have investigated the problem of how the electron 
density is modified when we change ${\cal E}_c/E_0$ for a fixed 
value of $a$. In the $x$-direction the charge-density is constant
while in the lateral $y$-direction it is modified because of the 
finite width of the system. 
Electron density at ${\bf r}$ is evaluated numerically from
$$ \rho({\bf r})=\sum_{i,j=1}^\infty\psi_i^*({\bf r})
               \psi_j({\bf r}) a_i^\dagger a_j.$$
Let us choose a particular value of $a$, say $a=8$, and see how 
the density profile of the translationally invariant state 
changes as a function of  ${\cal E}_c/E_0$. In Fig. 2 (a), 
we show the charge-densities as 
a function of ${\cal E}_c/E_0$. With increasing ${\cal E}_c/E_0$,
the width of the charge-density profile changes abruptly from one
value to the other. Calculating the effective filling factor from
$\nu=2\pi\lambda^2 n$, (where $n$ is the number of electrons per 
unit area) and taking the width as full width at half maximum, we
get the filling factors $0.98$, $0.71$, $0.56,\cdots,0.51$
and $0.42$ which are very close to the values $\nu=1$, $\frac23$,
$\frac12$ and $\nu=\frac25$. Repeating the same calculation at 
$a=9.5$ we get the densities shown in Fig. 2 (b). The effective 
filling factors for this value of $a$ are, $0.99$ and 
$0.68,\cdots,0.66$ which suggest that these states are $\nu=1$ 
and $\nu=\frac23$, respectively. The state which has the 
effective filling factor $0.38,\cdots,0.37$ is identified as a 
$\nu=\frac13$ state by the overlap calculation.

In Fig. 3 we show a phase diagram for the 1D-FQHE states. The 
diagram is obtained by systematically seeking those points in
the parameter space spanned by $a$ and ${\cal E}_c/E_0$ where the
ground state has {\it zero total momentum}. What is then plotted 
is the energy gap between this ground state and the first excited
state. In the figure the area of a filled dot is proportional to
that gap. The phase diagram consists of separate regions of 
several FQHE states. The filling factors of these states are 
marked on the figure. Remarkably, there is a distinct region for 
the {\it even-denominator} state $\nu=\frac12$. The area of this 
region is of course, much smaller than those with odd-denominator
states. But, given the total absence of the $\frac12$-state in a
single-layer system, this observation is rather unique. Fig. 4
depicts the energy spectra for the states $\nu=\frac23$,
$\frac12$, $\frac25$ and $\nu=\frac13$. These states are chosen 
from the phase diagram (Fig. 3) at the points where the gap 
appears to be the largest. The ubiquitous incompressible gaps in 
the spectra makes the analogy with those in the corresponding 
two-dimensional systems quite obvious. The novel result here 
again is, of course, the signature of incompressibility in the 
energy spectrum for the lowest half-filled Landau level, which is
a hallmark of the fractional quantum Hall state at that filling 
factor.

Between the stable 1D-FQHE regions in the phase diagram, the
system is in states with $M\ne0$. This suggests that the
symmetry changes between ground states in different regions of
the phase diagram. One possible broken symmetry state might be a 
phase in which the density is no longer symmetric around $y=0$, 
{\it i.e.}, some of the density is displaced from left side of 
the channel to the right side (or vice versa) spontaneously
\cite{Halperin}. Interestingly, for values of $a$ where broken
symmetry states appear, the minimum value of the excitation gap
seems to collapse towards zero \cite{future}. However, there can 
never be a true long-range order in the spacing of electrons 
along the wire (in the thermodynamic limit). In fact, in a 
one-dimensional system, no sharp distinction exists between a 
Wigner crystal with quantum fluctuations and a spinless Luttinger
liquid. In either case the density-correlation function has
a power-law singularity at a wavevector equal to twice the 
Fermi wavevector, and the exponent of the singularity varies 
continuously with the interaction parameters. These issues will 
be discussed elsewhere \cite{future}.

In conclusion, we have investigated the properties of a system of
electrons interacting via the long-range Coulomb interaction
in a narrow channel and in the quantum Hall regime. As
the interaction strength is increased, we notice abrupt jumps in 
the expectation values of the kinetic and potential energies of
translationally invariant states. The
width of the charge-density profile also revealed similar abrupt 
changes. We have calculated the phase diagram of the stable 
1D-FQHE states. In addition to various odd-denominator filling 
factors which are well established in the two-dimensional
systems, we find in a region of the parameter space the lowest
half-filled Landau level also appear as a stable incompressible
state. We also present the energy spectra of those incompressible
states. The low-lying collective modes at $\nu=\frac12$ are 
strikingly similar to those of an odd-denominator FQHE state. 

One of us (TC) would like to gratefully acknowledge very helpful 
discussions with B. I. Halperin (Harvard).

\begin{figure}
\caption{Expectation value of (a) kinetic energy per particle
and (b) interaction energy per particle, as a function of 
${\cal E}_c/E_0$ and length of the cell for the state $M=0$. The 
effective filling factors $\nu=\frac13,\frac23,\frac25,\frac27,$ 
and $\nu=1$ are also indicated.}
\label{fig1}
\end{figure}

\begin{figure}
\caption{(a) Electronic densities in the lateral direction
at $a=8$ and for $M=0$ states. (b) Similar results for 
$a=9.5$. The effective filling factors are shown in the figure.}
\label{fig2}
\end{figure}

\begin{figure}
\caption{Phase diagram for the FQHE states at the effective
filling factors $\nu=\frac13,\frac23,\frac25,\frac12,\frac27$
indicated in the figure.}
\label{fig3}
\end{figure}

\begin{figure}
\caption{Energy spectra calculated at $(a, {\cal E}_c/E_0)$ and 
$\nu$: (a) $(6.8, 24), \nu=\frac23$, (b) $(7.6, 36), 
\nu=\frac12$, (c) $(9.2, 40), \nu=\frac25$ and (d) $(12.2, 42), 
\nu=\frac13$. 
}
\label{fig4}
\end{figure}

\begin{references}
\bibitem{fqhe}
D. C. Tsui, H. L. St\"ormer, and A. C. Gossard, Phys. Rev. Lett.
{\bf 48}, 1559 (1982); H. L. St\"ormer, Physica B {\bf 177}, 401
(1992).
\bibitem{laughlin}
R. B. Laughlin, Phys. Rev. Lett. {\bf 50}, 1395 (1983).
\bibitem{halperin}
B. I. Halperin, Helv. Phys. Acta {\bf 56}, 75 (1983).
\bibitem{book}
T. Chakraborty and P. Pietil\"ainen, {\it The Quantum Hall
Effects} (Springer, New York, 1995), second edition; T.
Chakraborty, in {\it Handbook on Semiconductors}, vol. 1, ed.
P. T. Landsberg (North-Holland, Amsterdam, 1992), Ch. 17.
\bibitem{expt}
H. W. Jiang et al., Phys. Rev. B {\bf 40}, 12013 (1989); R. L. 
Willett, et al., Phys. Rev. Lett. {\bf 65}, 112 (1990); W. Kang 
et al., {\it ibid.} {\bf 71}, 3850 (1993); R. R. Du et al., {\it 
ibid.} {\bf 70}, 2944 (1993); {\it ibid.} {\bf 73}, 3274 (1994); 
D. R. Leadley, et al., {\it ibid.} {\bf 72}, 1906 (1994);
R. L. Willett et al., {\it ibid.} {\bf 75}, 2988 (1995).
\bibitem{theory}
B. I. Halperin, P. A. Lee, and N. Read, Phys. Rev. B {\bf 47}, 
7312 (1993); B. I. Halperin, Surf. Sci. {\bf 305}, 1 (1994).
\bibitem{electron}
F. D. M. Haldane, Phys. Rev. Lett. {\bf 55}, 2095 (1985);
G. Fano, F. Ortolani, and E. Tosatti, Nuovo Cimento {\bf 9}D,
1337 (1987); T. Chakraborty and P. Pietil\"ainen, Phys. Rev. B 
{\bf 38}, 10 097 (1988); M. Greiter, X. -G. Wen, and F. Wilczek, 
Nucl. Phys. B {\bf 374}, 567 (1992).
\bibitem{okiji}
A. Okiji and N. Kawakami, (Eds.), {\it Correlation Effects in 
Low-Dimensional Electron Systems} (Springer, 1994). 
\bibitem{timp}
G. Timp, R. Behringer, J. E. Cunningham, and R. E. Howard, Phys. 
Rev. Lett. {\bf 63}, 2268 (1989); G. Timp, in {\it Nanostructured
Systems}, ed. M. Reed (Academic, Boston, 1992), Ch. 3.
\bibitem{chui}
S. T. Chui, Phys. Rev. Lett. {\bf 56}, 2395 (1986).
\bibitem{frad}
P. Jordan and E. P. Wigner, Z. Phys. {\bf 47}, 631 (1928).
\bibitem{greiter}
M. Greiter, X.-G. Wen, and F. Wilczek, Phys. Rev. B, {\bf 45}, 
9489 (1992); Phys. Rev. Lett. {\bf 66}, 3205 (1991).
\bibitem{yoshioka1}
D. Yoshioka, J. Phys. Soc. Jpn. {\bf 62}, 839 (1993).
\bibitem{yoshioka2}
D. Yoshioka, Physica B {\bf 184}, 86 (1993).
\bibitem{Halperin}
B. I. Halperin, private communications (1996).
\bibitem{future}
T. Chakraborty, K. Niemel\"a, and P. Pietil\"ainen, to be
published.
\end{references}
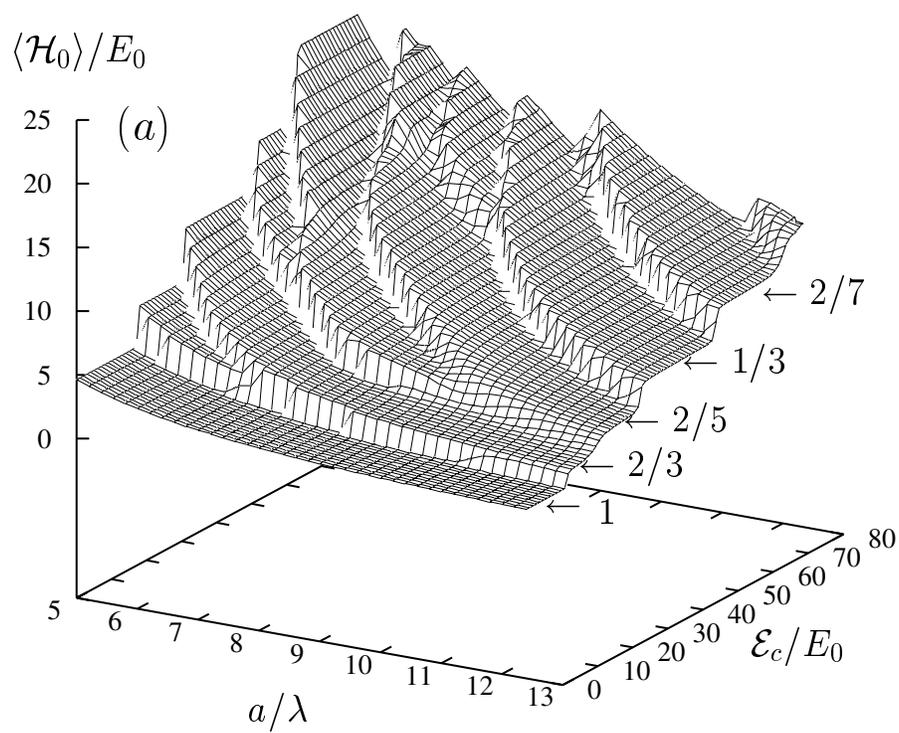
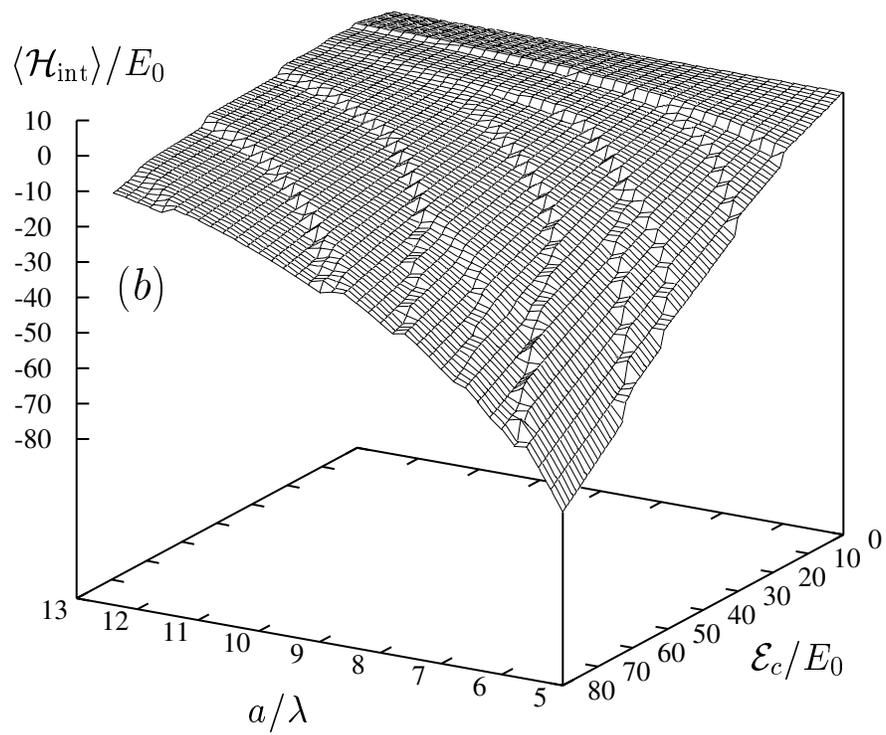
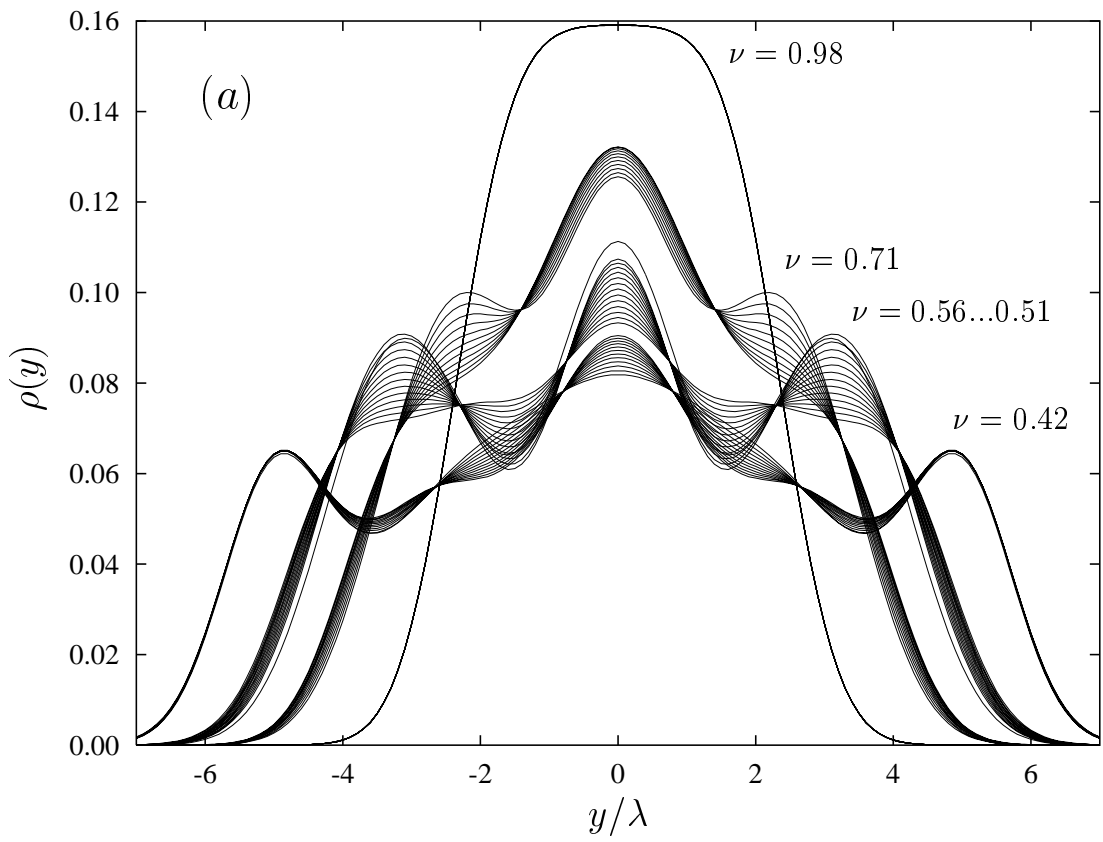
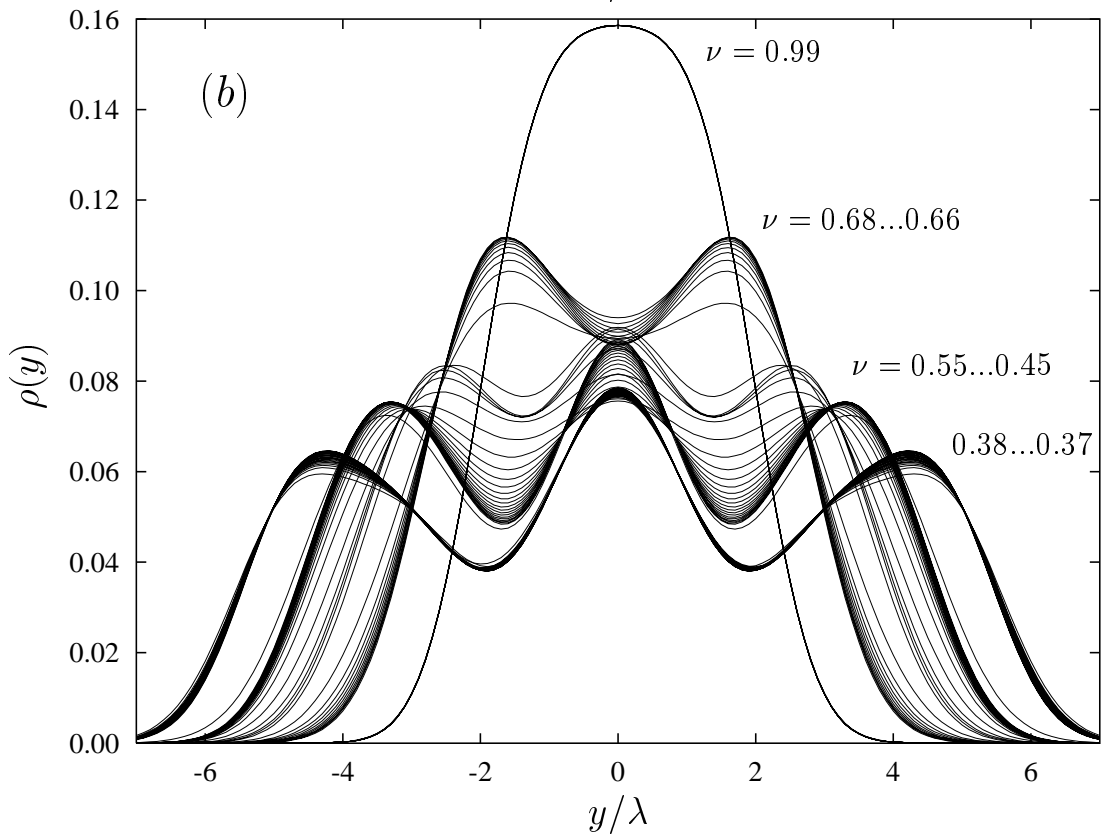
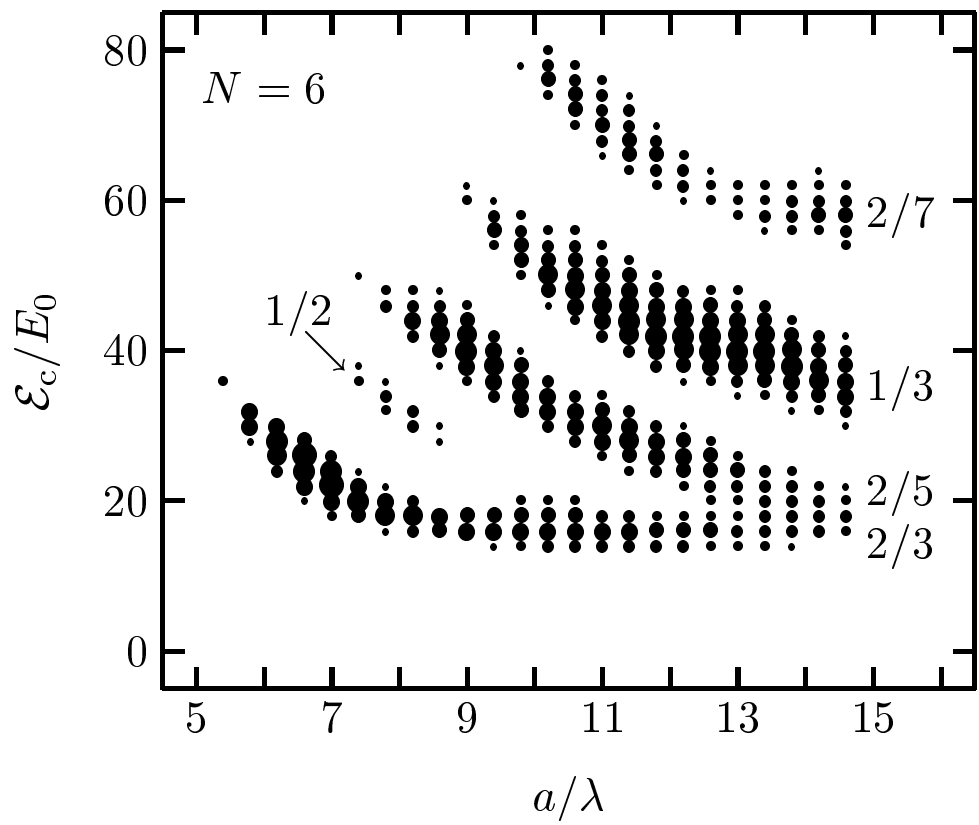
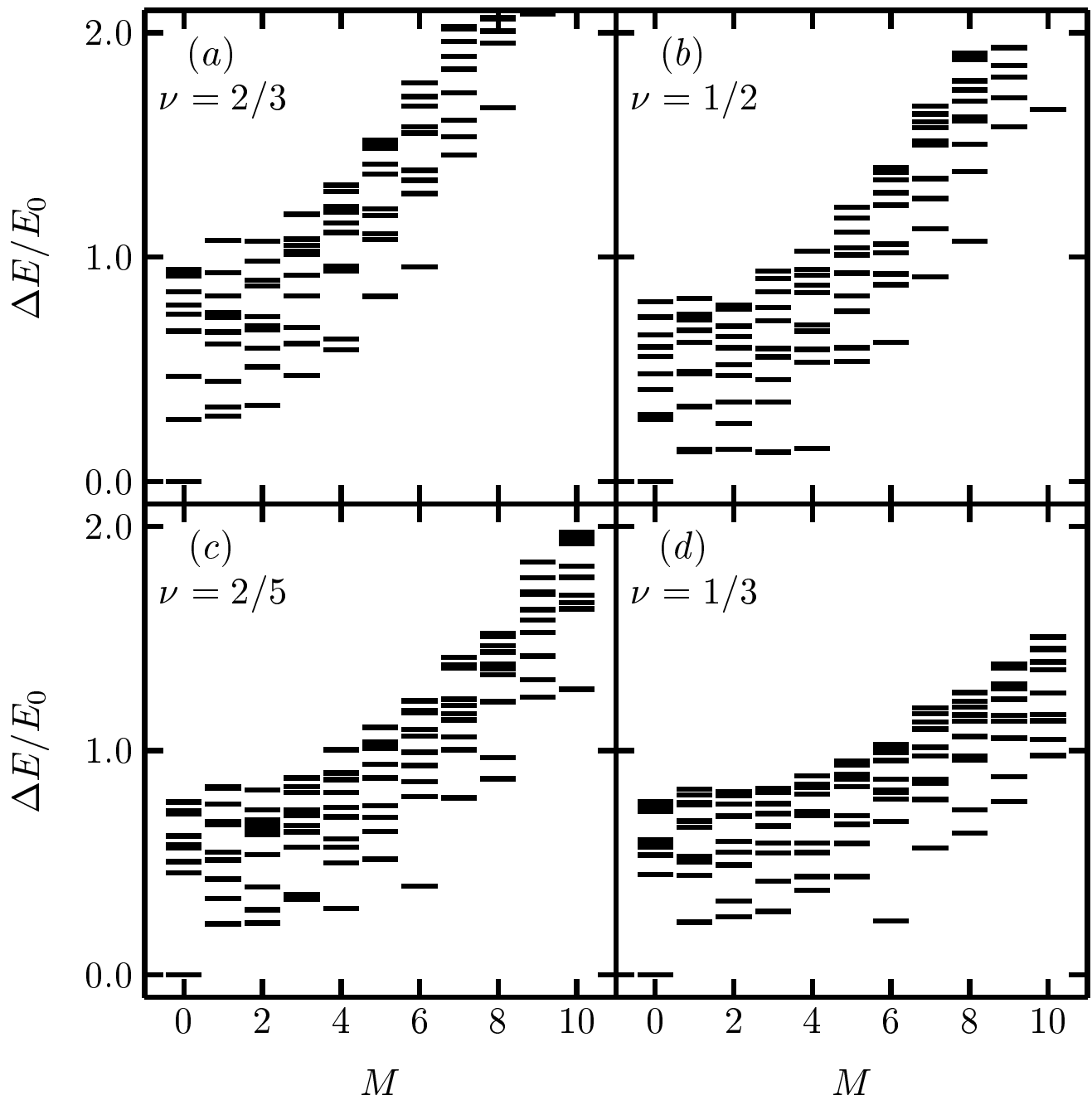
\end{document}